\documentclass[10pt,a4paper,twocolumn]{revtex4}
\usepackage{graphicx}
\usepackage{amsfonts}
\usepackage{amsbsy}
\newcommand{\beq}{\begin{equation}}
\newcommand{\eeq}[1]{\label{#1}\end{equation}}

\def\be{\begin{equation}}
\def\ee{\end{equation}}
\def\bea{\begin{eqnarray}}
\def\eea{\end{eqnarray}}

\begin{document}

\title{ Modified Boltzmann Transport Equation and Freeze Out}

\author{
L.P. Csernai$^{1,2}$, V.K. Magas$^3$, E. Moln\'ar$^1$,
A. Nyiri$^1$ and K. Tamosiunas$^1$\\}

\affiliation{
$^1$ Section for Theoretical and Computational Physics, and Bergen
Computational Physics Laboratory, BCCS-Unifob, University of Bergen,
Allegaten 55, 5007 Bergen, Norway\\
$^2$ MTA-KFKI, Research Inst of Particle and Nuclear Physics,
H-1525 Budapest 114, P.O.Box 49, Hungary\\
$^3$ Departamento de F\'{\i}sica Te\'orica and IFIC
 Centro Mixto\\ Universidad de Valencia-CSIC,
 Institutos de Investigaci\'on de Paterna\\ Apdo. correos 22085,
 46071, Valencia, Spain
 }

\begin{abstract}
{Abstract: We study Freeze Out process in high energy heavy ion reaction.
The description of the process is based on the Boltzmann Transport Equation (BTE). We point
out the basic limitations of the BTE approach and introduce Modified BTE.  The Freeze Out
dynamics is presented in the 4-dimensional space-time in a layer of finite thickness, and
we employ Modified BTE for the realistic Freeze Out description. }
\end{abstract}

\maketitle

\section{Introduction}
The Freeze Out (FO) is an important phase of dynamical reactions.
It is of primary importance in case of rapid, dynamical processes
where the originally strongly-interacting and locally equilibrated
matter undergoes a rapid explosive process, in which matter properties
change considerably, the interaction vanishes in a relatively
small space-time layer, and local equilibrium disappears.
The connection of the kinetic description of this process and
the Boltzmann Transport Equation (BTE) raised considerable attention recently
\cite{A,D}.

The problem is to calculate the phase-space (PS) distribution of the post FO
particles.  Earlier such kinetic FO calculations were performed in
one-dimensional models \cite{FO1,FO3-HIP,FO4,FO5}, where the dynamics was
governed by two constants: a re-thermalization parameter and a FO
parameter. This latter one is governed by the phase-space FO probability,
which was constructed recently in a fully covariant form \cite{Cs04}.

The FO is a kinetic process and one would think it can be
handled perfectly by using the Boltzmann Transport Equation, which
may describe equilibrium and non-equilibrium processes equally
well in a 4-dimensional space-time volume element, which is usually
a FO layer. This work and ref.
\cite{D} follows this approach. This finite layer is frequently
idealized as a 3-dimensional FO hypersurface. In ref.
\cite{A,B} author analyzes the features of this idealized
discontinuity
\footnote{
In refs. \cite{A,B} the author discusses two physically different
situations, which should be clearly separated,
otherwise one can run into some confusions. The first
one is the "transition" from hydrodynamical description
of the system created in heavy ion collisions to cascade
description. This is not a physical phase
transition, just a switch from one theoretical model
to another, which can be justified, strictly speaking,
only in the region where both models give adequate
description of the system, i.e. the same result for the all
possible observables. In this overlapping ST region such
a transition can be realized at any infinitely
narrow dividing hypersurface. On the other hand, for the
case of a real physical phase transition, like FO
or/and hadronization, the infinitely narrow
dividing hypersurface is an idealization of the
layer of some finite thickness, as it was discussed
above. We basically assume that if we use the correct
particle distributions on the "post" side, we do not
make a big mistake changing distributions to the new
ones sharply instead of changing them gradually in some
layer.}.

The FO can also be simultaneous with a phase transition, especially
when the phase transition reduces the number of degrees of freedom
and contributes to the FO process this way.
As an example let us describe a gradual hadronization and FO
of the Quark-Gluon Plasma in a  layer, where quasi-hadrons
 or hadrons are formed, the new particles gradually cease to
interact, their PS distribution changes and the matter
gradually freezes out.

Free hadrons, which are formed, do not interact
with anything and propagate directly to the detector. Although, the
formation of these fragments can be most suitably described
in a coalescence or recombination model, most finally observed
baryon abundances follow the statistical model predictions. The reason is
simple: the formation cross sections are governed by the same statistical
factors as the thermal equilibrium, because the radial part of the
formation probability for s-wave hadrons is about the same. Exceptions are
the excited states, e.g. the p-wave hadrons like $\Lambda (1520)$,
which have a smaller radial form factor and consequently they are suppressed
in comparison to the statistical equilibrium abundance, which is sensitive
to their weight only.

In this work we do not discuss issues related to simultaneous hadronization
and FO, for simplicity we consider one type of particles only
and study their kinetic evolution. If hadronization
happens simultaneously with kinetic FO the
kinetic description presented here can and should be extended. For example,
some features of the fast hadronization and FO of supercooled Quark-Gluon Plasma, which
might be created in ultra-relativistic heavy ion collisions (for the first time such a
scenario was proposed in refs. \cite{CC94,CM95})
are discussed in refs. \cite{KCM03,CAK03}.
The simultaneous
FO and hadronization can be described in an idealized way
by using the 3-dimensional FO hypersurface approach as suggested
in ref. \cite{A}. Then this simplified approach makes it possible
to solve some of the basic problems. The simultaneous hadronization
and FO can also be handled by assuming idealized hadronization
at the hypersurface, which is the inside boundary of the FO
layer of finite thickness, $L$. In this way the present work is
also relevant for the simultaneous hadronization and FO problem.

In the present work we analyze the situation, discuss the
applicability of BTE, and point out the physical causes which,
limit the applicability of the BTE for describing FO. And the
aim is to show how can we overcome this obstacle. For this purpose we will modify the
BTE and then will show how one can derive out of it a simple one dimensional
kinetic model, similar to the one used by some of the authors in earlier calculations.

\section{Particles emerging from Freeze Out hypersurface}

Not only in heavy ion reactions, but in many dynamical processes
particle creation (or condensation) happens mostly in a directed way: the
phenomenon propagates into some direction, i.e. it happens in
some layer or front (like detonations, deflagrations, shocks,
condensation waves or FO across a layer with space-like normal).
The reason is that neighboring
regions in the front may interact to minimize the energy of the front
by evening it out, providing energy to neighboring regions to
exceed the threshold conditions. Even in those relativistic processes that
are time-like (have time-like normal), and so the neighboring points
of a front cannot be in causal connection,
the dynamical processes may and frequently have a direction. See the example
in ref. \cite{Cs87}. This can be a simple consequence of the
initial and boundary conditions.

These fronts have a characteristic direction (or normal, $d\sigma^\mu$).
Let us look at an example when particles in a domain of
the space-time (ST) are characterized by a
phase-space distribution, $f(x,p)$. Then the space-time current density
of these particles, $N^\mu(x)$ can be described as
\be
    N^\mu = \int  \ \frac{d^3p}{p^0}\ p^\mu \ f(x,p)\ .
    \label{eq01}
\ee
The net number of particles crossing an arbitrary hypersurface element
$d \sigma_\mu$ is
\be
  dS = N^\mu  d \sigma_\mu =
   \int \frac{d^3p}{p^0}\ p^\mu\; d \sigma_\mu\ f(x,p)\ .
    \label{eq02}
\ee
If we want to describe the FO, particles are allowed to cross the FO
hypersurface "outwards" only,
i.e., only in the direction of  $d\sigma^\mu$. Thus,
\bea
   S_{FO} &=& \int N_{FO}^\mu  d \sigma_\mu  \\ \nonumber
     &=& \int  \int  \ \frac{d^3p}{p^0}\ p^\mu\; d \sigma_\mu\ f_{FO}(x,p)
       \Theta(p^\mu\; d \sigma_\mu)\ ,
    \label{eq03}
\eea
where either the phase space distribution, $f_{FO}(x,p)$, should have only
particles with momenta pointing outwards (post FO distribution), and/or
this is secured by the step function $ \Theta(p^\mu\; d \sigma_\mu)$.
Eq. (\ref{eq03}) yields the modified Cooper-Frye FO formula, where
$f_{FO}(x,p)$ should be determined in such a way that all conservation
laws across the FO hypersurface are satisfied and overall entropy does
not decrease! \cite{CF74,Bu96,FO1}

\section{Non-isotropic particle sources}

\begin{center}
\begin{figure}[!ht]
%\vspace{-0.8cm}
    \includegraphics[height=8.5cm, width = 6.5cm, angle=270]{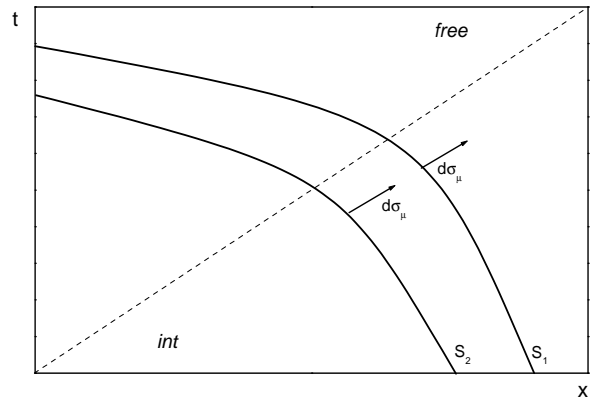}
\vspace{-0.2cm}
 \caption{\label{fig_layer}
Space-time picture of the FO process. At early times, centrally in the
collision region we have intensively interacting matter, which is
equilibrated and thermalized, this is the fluid-dynamical domain bordered
by the $S_2$ hypersurface, which has a normal 4-vector $d\sigma_\mu$.
The location of this surface is given by the fact
that the interacting fluid is cooling and expanding, and reaches a point when
interactions are not frequent enough to maintain full thermal and hydrodynamical
equilibrium locally. Some particles will not interact any more beyond this
hypersurface. Later on in the expansion and cooling we reach an other
hypersurface, $S_1$. By reaching this surface on their way all
particles become non-interacting, or free. Thus, when reaching this surface
the FO process is completed. The momentum distribution
of particles does not change any more. This is the (post) FO
distribution}
\vspace{-0.6cm}
\end{figure}
\end{center}

The FO-fronts or FO-layers are not necessarily narrow, but they have
a characteristic direction (or normal, $d\sigma^\mu$), and it is more
realistic to assume a continuous, 4-volume FO in a layer (or domain)
of the space-time. At the inside boundary of this layer
no particles are frozen out yet, while at the outside boundary hypersurface
all particles are frozen out and no interacting particles remain
(see Figure \ref{fig_layer}). For the sake of simplicity let us also assume that the
total particle number is conserved, even if simultaneous freeze out,
hadronization and particle formation are frequently discussed.

Thus, while the total number of particles remains constant, in this
domain, the number of interacting particles decreases and the
number of frozen out or free particles increases:
\bea
       N^\mu(x) & = &  N_i^\mu(x)  +   N_f^\mu(x) \ , \\
       \partial_\mu N^\mu(x)     & = & 0         \\
       \partial_\mu N_i^\mu(x)   & = & - \ \partial_\mu N_f^\mu(x) \ .
\eea

Then the space-time (ST) volume element, $d^4x$, in the layer of interest
can be converted into
$
 d^4 x  \longrightarrow  ds^\mu \  d\sigma_\mu ,
$
where $ds^\mu$ is the length element in the direction of the 4-vector
$d\sigma^\mu$ , which can be space-like or time-like, i.e.:
time-like,
$d\sigma^\mu \ d\sigma_\mu = + 1$,
or space-like,
$d\sigma^\mu \ d\sigma_\mu = - 1$.

%----------------------------figure 1 ---------------------------------------
%---------------------------end figure ----------------------------------------

Let us also assume that the ST domain,
where free particle formation happens,
is a layer, which is relatively narrow compared to the
bulk of matter (see Figure \ref{fig_layer}).
Assume also that the boundaries of this layer are
parallel or approximately parallel, and the thickness of the
layer does not vary much. Under these conditions one can
describe the change of free particle number in the layer
via the divergence of the 4-current of the particles
by the expression:
\bea
   \Delta N_i &=&
   \int d^4x\ \
   \partial_\mu N_i^\mu(x) \\ \nonumber
&=&
   \int ds^\mu \  d\sigma_\mu
   \int d^3p \ \frac{p^\mu}{p^0}\ \partial_\mu
         f_i(x,p)\,.
    \label{eq11}
\eea
According to the physical assumptions discussed
above, the
4-divergence is maximal in the direction of $d\sigma_\nu$, and
negligible in the other 3 orthogonal directions.

The emission or freeze out  probability may depend on physical
processes, cross sections, transition rates, and the actual
PS distributions.   Furthermore, $f_{i}$ can be space-time dependent,
and must be determined {\it self-consistently} during the detonation,
deflagration or FO process \cite{FO1,FO2,FO3-HIP,FO4,FO5}.

We will return to realistic FO probabilities later in section \ref{escape}.

\section{Freeze Out and the Boltzmann Transport Equation}
\label{secBTE}

One can derive the Boltzmann Transport Equation from the conservation
of charges in a ST domain \cite{LasBook}, $\Delta^4x$, assuming the
standard conditions:
(i) only binary collisions are considered,
(ii) we assume "molecular chaos", i.e. that the number of
        binary collisions at position $x$ is proportional to
        $f(x,p_1) \times f(x,p_2)$, and that
(iii) $f(x,p)$ is a smoothly varying function compared to the
mean free path (m.f.p.). The conservation
laws lead then to the requirement that the integral of the 4-divergence of
conserved charges should vanish
\beq
\int_{\Delta^4x} \int_{\Delta^3p}d^4x\frac{d^3p}{p^0}p^\mu
\partial_\mu f(x,p) = 0 \ .
\eeq{int}\\
As the choice of the ST 4-volume element is arbitrary we obtain
the differential form of the conservation law, which describes the
evolution of the PS distribution, $f(x,p)$, of a particle with momentum
$p$. However, if we
take into account that particles can scatter into this PS volume
element around $p$, or can scatter out from this volume element, we have
to add Gain- and Loss- collision terms to the conservation equation
(see e.g. sect 3.2 of ref. \cite{LasBook}):
\bea\label{BTE1}  \nonumber
p^\mu \partial_\mu f(p)
&=&
 \frac{1}{2} \int {_{12}\mathcal{D}}_4    f(p_1)f(p_2) W_{p_1p_2}^{pp_4} \\
&-& \frac{1}{2} \int {   _2\mathcal{D}}_{34} f(p)  f(p_2) W_{pp_2}^{p_3p_4}\ .
\eea
Here we assume elementary collisions where in the initial state
two particles collide with momenta
$p_1$ and $p_2$
into a final state of two particles with momenta
$p_3$ and $p_4$. In case of the Gain term the particle described by the
BTE, with momentum $p$ (without an index), is one from the two final
state particles, while in case of the Loss term this particle is
one of the initial state particles. This is indicated by the
indexes of the invariant transition rate \cite{LasBook}.
We integrate over the momenta of the other three particles
participating in this binary collision. We use the notation
$$
{_{12}\mathcal{D}}_3 \equiv
\frac{d^3 p_1}{p_1^0}
\frac{d^3 p_2}{p_2^0}
\frac{d^3 p_3}{p_3^0} \ .
$$
We can shorten the notation further by suppressing the arguments
of the PS distribution functions, and the indexes of the
momenta in the argument will be carried by the distribution function
$f$ and the collision term $W_{p_1p_2}^{pp_4} \equiv W_{12}^{p4}$:
\beq
p^\mu \partial_\mu f=
 \frac{1}{2}\int {_{12}\mathcal{D}}_4    f^i_1 f^i_2 W_{12}^{p4}
-\frac{1}{2}\int {   _2\mathcal{D}}_{34} f^i   f^i_2 W_{p2}^{34}
\eeq{BTE2}
Now, aiming for the description of FO process let us
split up the distribution function, $f$, into
$f=f^i+f^f$, where $f^f$ is the phase-space distribution function
of the "free" or frozen out particles, which do not collide any more,
while $f^i$ is the interacting component \cite{FO1,D}.
Thus, the FO process is represented here by gradually
populating and building up the "free" component,
while draining particles from the interacting component.
As the particles belonging to the free component may not collide any more,
they do not appear in the initial state components of collision integrals!
\beq
p^\mu \partial_\mu (f^i+f^f)=
 \frac{1}{2} \int {_{12}\mathcal{D}}_4    f^i_1f^i_2 W_{12}^{p4}
-\frac{1}{2} \int {   _2\mathcal{D}}_{34} f^i  f^i_2 W_{p2}^{34}\ .
\eeq{BTE3}
The gain term, $ f^i_1\ f^i_2\ W_{12}^{p4}$ populates both
the interacting, $f^i$, and free, $f^f$, components,
so we will introduce a FO probability, which 'feeds'
the free component. The probability is phase-space dependent.
In principle it may depend on the positions and momenta of both
incoming particles, and it can weight the outgoing phase space
for one (or both) outgoing particles. In the most simple case
we have to assume that it depends at least on the momentum of the
outgoing particle, which belongs to the component $f^f$:
${\mathcal{P}}^{FO}(x,p)\equiv {\mathcal{P}}_f$.
\bea\nonumber
p^\mu \partial_\mu (f^i+f^f)&=&
 \frac{1}{2} \int {  _{12}\mathcal{D}}_4    f^i_1 f^i_2
                \Big[    {\mathcal{P}}_f  W_{12}^{p4}+
                     (1-{\mathcal{P}}_f) W_{12}^{p4}\Big] \\
&-&\frac{1}{2} \int      {_2\mathcal{D}}_{34} f^i   f^i_2
                                          W_{p2}^{34}.
\eea\label{BTE4}
Now, we can separate the two components into two equations.
The sum of these two equations returns the complete BTE above:
\beq
p^\mu \partial_\mu f^f=
 \frac{1}{2}\int {_{12}\mathcal{D}}_4 f^i_1 f^i_2 \ {\mathcal{P}}_f W_{12}^{p4}
\eeq{BTEhalf1}
\beq
p^\mu \partial_\mu f^i=
 \frac{1}{2}\int {_{12}\mathcal{D}}_4    f^i_1 f^i_2 \
                                       (1-{\mathcal{P}}_f) W_{12}^{p4}
-\frac{1}{2}\int    {_2\mathcal{D}}_{34} f^i   f^i_2     W_{p2}^{34}
\eeq{BTEhalf2}
The free component does not have a Loss term, because particles in the
free component cannot collide, and so, the free component cannot loose
particles due to collisions.
Rewriting the second equation  yields:
\bea \nonumber
p^\mu \partial_\mu f^i  &=&
-\frac{1}{2}\int {_{12}\mathcal{D}}_4    f^i_1 f^i_2 {\mathcal{P}}_f W_{12}^{p4}
+ \frac{1}{2}\int {_{12}\mathcal{D}}_4    f^i_1 f^i_2 W_{12}^{p4} \\
&-& \frac{1}{2}\int {   _2\mathcal{D}}_{34} f^i   f^i_2 W_{p2}^{34}
\eea\label{BTEhalf2a}
The first term is a drain term, describing the "escape" or "freeze out"
of particles from the interacting component.
It is the inverse of the gain term (or source term) for the free
component, $f^f$.
The last two terms are influencing the interacting term by redistributing
particles in the momentum space. These latter two terms do not include the
FO probability factors! Thus,
these two terms drive the interacting component towards re-thermalization.
As a usual approximation these two terms can be approximated
by the relaxation time approximation as in refs. \cite{FO3-HIP,FO4,FO5}.
Thus, the BTE describing FO in this situation reads as:
\beq
p^\mu \partial_\mu f^f=
\frac{1}{2}\int {_{12}\mathcal{D}}_4  f^i_1 f^i_2 {\mathcal{P}}_f W_{12}^{p4}
\eeq{BTEhalf1b}
\beq
p^\mu \partial_\mu f^i=
-\frac{1}{2}\int {_{12}\mathcal{D}}_4 f^i_1 f^i_2 {\mathcal{P}}_f W_{12}^{p4}
\ +\ p^0\ \frac{f^i_{eq}-f^i}{\tau_{rel}}
\eeq{BTEhalf2b}
The first equation, eq. (\ref{BTEhalf1b}), describes the gain of the
free component, i.e. that part of the earlier gain term, which
will not collide any more.
The first term in the second equation has the same value with opposite sign.
This describes the part of
$f^i$, which is leaving the interacting component and does not take part
in the re-thermalization.

In fact the above described collision integrals can be highly simplified, by
exploiting the symmetries and conservation laws in the invariant
transition rate, $W$, so that
only one phase-space integral remains to be executed
(see section 3.3 and eq. (3.27) in ref. \cite{LasBook}).

\section{Modified Boltzmann Transport Equation}
\label{secMBTE}

Now, the question arises: can the BTE handle realistically the
FO process? We have seen that the structure of the kinetic equations,
used earlier to describe FO \cite{FO1,FO3-HIP,FO4,FO5}, and the
separation of the "escape" and "re-thermalization" terms come
out in a simple, straightforward way from the BTE.

However, the usual structure of the collision terms in the
BTE are not adequate for describing rapid FO, in a layer which is
comparable to the m.f.p. If we assume the existence of such a
layer this immediately
contradicts assumption (iii):
the change is not negligible in the direction of
$d\sigma^\nu$. The assumption (ii) of
"molecular chaos" is also violated in a FO process because
number of collisions is not proportional with
$f(x,p_1) \times f(x,p_2)$, but it is delocalized
in the normal direction with
$f(x+\lambda,p_1) \times f(x-\lambda,p_2)$.
(The fact that the FO is a delocalized kinetic process, was
already used in ref. \cite{D} when integrals along the
path of propagating particles were introduced, but the consequences regarding
the details of the collision terms and the validity of the molecular chaos
assumption were not discussed.)

Based on the above considerations, one might conclude that the changes
of the distribution function are mediated by the transfer of particles, and
consequently only slowly propagating changes are possible. I.e., the front
propagates slowly, and its normal, $d\sigma^\mu$, is always space-like.
This was a common misconception, until recently, where all "superluminous"
shock, detonation, deflagration fronts or discontinuities were considered
unphysical based on early studies \cite{Taub}.
However, it was shown recently, that discontinuous changes may
happen simultaneously in spatially neighboring points, i.e. the
normal of the discontinuity-hypersurface can be time-like \cite{Cs87,C}.
This applies to the FO process also. Thus, the direction of characteristic
or dominant change,  $d\sigma^\mu$, may be both space-like and time-like
in the FO process.

From all the processes mentioned above (i.e. shocks, detonations, deflagrations etc.)
the FO is the most special one. Because the number of interacting particles is constantly
decreasing as the FO proceeds and correspondingly the m.f.p. is increasing and, in
fact, it reaches infinity
when the complete FO is finished. This simply means that we strictly speaking
can not make FO in finite layer of
any thickness smooth enough to be modeled with BTE. It is also obvious that if FO has some
characteristic length scale (thickness of the layer or even some characteristic parameter for infinitely
long FO \cite{FO3-HIP}), it is not proportional with the m.f.p., because m.f.p. increases as the density of
interacting component becomes smaller, while FO becomes faster in this limit, so its
characteristic scale should decrease.

Since, there is
a strong gradient in the FO direction: the free component
rapidly increases, while the interacting component decreases
along the FO direction, we can conclude that the
collision terms in their usual form are not adequate
to describe the FO process, particularly not the
"escape" probability or "escape" term.  The appropriate equations
to describe this system can be a Modified Boltzmann Transport Equation (MBTE) \cite{praga04} :
\bea\label{mbte}  \nonumber
p^\mu \partial_\mu f(p)
&=&
 \frac{1}{2} \int {_{12}\mathcal{D}}_4    \overline{f(x,p_1)}^{\ x}\ \overline{f(x,p_2)}^{\ x} W_{p_1p_2}^{pp_4} \\
&-& \frac{1}{2} \int {   _2\mathcal{D}}_{34} \overline{f(x,p)}^{\ x}\ \overline{f(x,p_2)}^{\ x} W_{pp_2}^{p_3p_4}\ ,
\eea
where $\overline{f(x,p_i)}^{\ x}$ is an average over all possible origins of the particle in the
backward lightcone of
the ST point $x=(t,\vec{x})$:
\begin{widetext}
\beq
 \overline{f(x,p)}^{\ x}=\frac{\int_{t_0}^{t} dt_1 \int d^3x_1 \delta^3(\vec{x} - \vec{x}_1 - \vec{v} (t-t_1)) f(x_1,p)
 e^{-\int_{t_1}^{t} dt_2 \int d^3x_2 \sigma n(x_2) v \delta^3(\vec{x}_2 - \vec{x}_1 - \vec{v} (t_2-t_1))}}
 {\int_{t_0}^{t} dt_1 \int d^3x_1 \delta^3(\vec{x} - \vec{x}_1 - \vec{v} (t-t_1))
 e^{-\int_{t_1}^{t} dt_2 \int d^3x_2 \sigma n(x_2) v \delta^3(\vec{x}_2 - \vec{x}_1 - \vec{v} (t_2-t_1))}},
\eeq{aver}
\end{widetext}
where $\delta^3(\vec{x} - \vec{x}_1 - \vec{v} (t-t_1))$ fixes the ST
trajectory, along which the particles with given momentum can reach the ST point $x$, time
$t_0$ is given by the
initial or boundary conditions,
$\vec{v} = \vec{p} / p^0$ ($v=|\vec{v}|$),
and the exponential factor accounts for the probability not to have any
 other collision from the origin
$x_1$ till $x$. In the arguments of exponents $n(x)$ is the particle density in the calculational frame,
$n(x)=N^0(x)$, and
$\sigma$ is the total  scattering cross section. After performing integrations over $d^3x$ with a help of
$\delta$-functions we can
write the MBTE equation in the form:
\bea\label{mbte2}  \nonumber
& & \hfill p^\mu \partial_\mu f(p) = \\
& = & \frac{1}{2} \int {_{12}\mathcal{D}}_4^{t_1t_2}    f(t_1,p_1) G(t_1,p_1)
f(t_2,p_2)  G(t_2,p_2) W_{p_1p_2}^{pp_4} \nonumber \\
&-& \frac{1}{2} \int {   _2\mathcal{D}}_{34}^{t_1t_2} f(t_1,p) G(t_1,p)
f(t_2,p_2)  G(t_2,p_2) W_{pp_2}^{p_3p_4}\,, \nonumber\\
& &
\eea
where
\beq
{_{12}\mathcal{D}}_4^{t_1t_2}=\frac{1}{2}\int_{t_0}^{t} dt_1 \int_{t_0}^{t} dt_2
\int {_{12}\mathcal{D}}_4\,,
\eeq{newnot1}
\beq
f(t_1,p)=f(t_1,\vec{x} - \vec{v} (t-t_1),p)\,,
\eeq{newnot1A}
\beq
G(t_1,p)=\frac{e^{-\int_{t_1}^{t} dt_2  \sigma n(t_2,\vec{x} - \vec{v} (t-t_2)) v }}{C(x,p)}\,,
\eeq{newnot2}
\beq
C(x,p)=\int_{t_0}^{t} dt_1
 e^{-\int_{t_1}^{t} dt_2  \sigma n(t_2,\vec{x} - \vec{v} (t-t_2)) v } \,.
\eeq{newnot2A}

Interestingly, Molecular Dynamics models do not use the
local molecular chaos assumption, and follow the trajectories of the colliding particles instead. Thus
such models do actually solve the MBTE, and not BTE, although this was not realized before.

The obvious limit in which  MBTE is reduced to BTE
is a completely homogeneous ST distribution function
(i.e. no external forces, no boundaries). Another possibility is
the hydrodynamic limit, $\lambda=1/\sigma n\rightarrow 0$, when the exponential factors
(\ref{newnot2},\ref{newnot2A})
will be reduced to $\sim \delta(t-t_{1,2})$,
reproducing the BTE after $t_1,t_2$ integrations.

The symmetries and the assumption of local
molecular chaos lead to the consequence that local conservation
laws can be derived from the original BTE, i.e.
$\partial_\mu T^{\mu\nu} = 0$ and
$\partial_\mu N^\mu = 0$, where $T$ and $N$ are given as integrals over the
single particle PS distribution and the momentum.
Although now we have delocalized the equations, the local conservation laws can still  be
derived in the same way, as it was shown in Ref. \cite{praga04}.
The very essential property of the BTE is the Boltzmann H-theorem. Here the situation is more complicated and the behaviour of the entropy current in MBTE is a subject of future studies.  Nevertheless,
for adiabatic expansion
($S^{\mu}_{,\mu}= 0$) a sufficient condition is the
same as for BTE, namely
$f(x,p_1)f(x,p_2)=f(x,p_3)f(x,p_4)$ \cite{praga04}.

The obtained MBTE is considerably more complicated, than the
original BTE. In order to proceed let us make a further simplification assuming that all the particles arrive into  the collision point $x$ from one m.f.p. distance
\cite{toPL} (instead of allowing them to arrive from any distance with the corresponding probabilities, as it is done in eq. (\ref{aver})). This then leads to the following simplified equation, which nevertheless is still  adopted to the strongly non-homogeneous systems much better then the original BTE:
\bea\label{mbte0}  \nonumber
p^\mu \partial_\mu f(p)
&=&
 \frac{1}{2} \int {_{12}\mathcal{D}}_4    f(\tilde x_1,p_1) f(\tilde x_2,p_2) W_{p_1p_2}^{pp_4} \\
&-& \frac{1}{2} \int {   _2\mathcal{D}}_{34} f(\tilde x,p)f(\tilde x_2,p_2) W_{pp_2}^{p_3p_4}\,,
\eea
where $x_k$ is the origin of colliding particles, i.e. the ST point
where the colliding particles were colliding last,
$\tilde x_k = x - u_k \tau(x,\vec{v}_k/v_k)$,
$u_k^\mu = (\gamma_k, \gamma_k \vec{v}_k)$,
$\gamma = 1 / \sqrt{1-\vec{v}^2}$
and $\vec{v}_k = \vec{p}_k / p_k^0$.  Here $\tau$ is the collision time, such that $|\vec{v}|\tau(x,\vec{v}/v)=\lambda(x,\vec{v}/v)$.   Note that the
m.f.p. depends not only on the position, but also on the direction of the particle motion. This is an essential modification
if the PS distribution has a large gradient in the space-time. This gradient
defines a ST 4-vector characterizing the direction of the process,
$d\sigma^\mu$.
In ref. \cite{D} the direction $d\sigma^\mu$
is also introduced, however, it is not discussed why and it is
not connected to the delocalization of the BTE.

For the FO modeling, repeating for the eq. (\ref{mbte0}) the same step as for BTE above, we then obtain:
\beq
p^\mu \partial_\mu f^f(x,p) =
\frac{1}{2}\int {_{12}\mathcal{D}}_4   {\mathcal{P}}_f W_{12}^{p4}\
f^i(\tilde x_1,p_1) f^i(\tilde x_2,p_2) \ ,
\eeq{MBTE1}
\bea\label{MBTE2}\nonumber
p^\mu \partial_\mu f^i(x,p) &=&
-\frac{1}{2}\int {_{12}\mathcal{D}}_4  {\mathcal{P}}_f W_{12}^{p4}\
f^i(\tilde x_1,p_1) f^i(\tilde x_2,p_2) \\
& +& \  p^0\ \frac{f^i_{eq}-f^i}{\tau_{rel}} \ .
\eea

A simple general solution of the MBTE (\ref{MBTE1},\ref{MBTE2}) cannot be
given but it serves as a basis for simplified, phenomenological kinetic
models describing the FO process.

%-----------------

\section{Approximate Kinetic Freeze Out Models}

In this section our goal is to present a schematic derivation of a
simple kinetic FO model used by some of the authors earlier. This
represents only one particular possibility and the general
MBTE equation can be solved or approximated in other ways also. The
approximation we present is one of the simplest possibilities, but
not necessarily the most realistic one.

If the ST distribution is non-uniform and the direction of steepest
gradient can be clearly identified, one may replace one (or
more) of the integrals over $d^3p_1$ (or $d^3p_2$) by space-time
integrals over the origins of the incoming particle(s), $d^4x'$,
requiring that the particle reaches the ST point, $x$, when needed.
This requirement determines $p^\mu$ for a given $x'^\mu$.
It is reasonable to assume that after converting some of the
integrals to ST integrals and performing them, we get
an effective FO term reflecting the properties of the
local PS distribution, transition rate, the ST
configuration (e.g. gradient of density change, and its direction)
and characteristics of the FO layer.

Let us return to the basic integral form
of the kinetic theory, eqs. (\ref{MBTE1},\ref{MBTE2}), and discuss the
FO probability. We will study equation  (\ref{MBTE1}) without
performing the integrals in a formal way, rather illustrating
the procedure giving a better insight into the problem.

\begin{center}
\begin{figure}[!ht]
\vspace{-0.2cm}
    \includegraphics[height=7cm, width = 8.4cm]{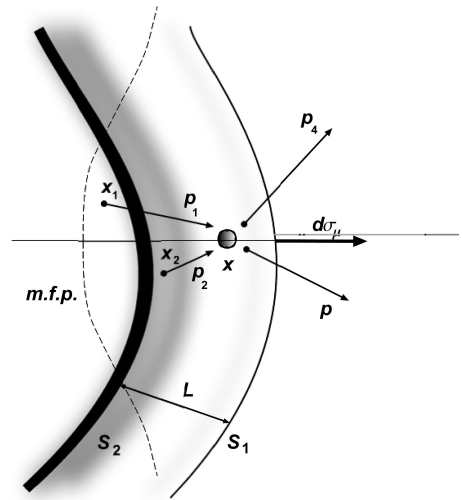}
\vspace{-0.2cm}
 \caption{\label{fig_mfp}
The plot of one of the  last collisions at $x$,
shown in the spatial cross section
of the FO layer. Particles arrive from positions $\tilde x_1$ and $\tilde x_2$ to point
$x$ with momenta $p_1$ and $p_2$. Within the FO layer of thickness, $L$,
the density of interacting particles gradually decreases (indicated by
shading) and disappears at the outside boundary, $S_1$ (thin line) of the
layer. R.h.s. from this boundary there are no interacting particles.
Particles can reach $x$ from a region closer than the mean free path
(m.f.p. indicated by the dashed line), but only from places where the
interacting particle density is still not zero, i.e. mostly from the left.
The inside boundary of the FO layer,  $S_2$ (thick line) indicates the
points where the FO starts. Left of this line there is only interacting
matter and the FO probability is assumed to be zero for collisions happening
in the interacting region.
}
\vspace{-0.6cm}
\end{figure}
\end{center}

When we are in the FO layer, close to the boundary of complete
FO  we have to calculate here the
collision rate.
According to the MBTE this depends on the PS distribution of the
incoming particles at their origins,
$ f^i(\tilde x_1,p_1) \ f^i(\tilde x_2,p_2)$.
Assume that the FO direction points in the direction of $d\sigma^\mu$, as it is shown in Figure \ref{fig_mfp}. On the right hand side
of the collision point the density of interacting particles is low or zero, while on
the left hand  side it is larger, closer to the pre FO value
(see Figure \ref{fig_mfp}).
%----------------------------figure 2 ---------------------------------------
%---------------------------end figure ----------------------------------------
It is more probable, that particles arrive to the collision point $x$ from the left side, because of the higher density of the interacting particles on the left.
 Consequently, most outgoing particles leave to the right.
Thus, the collision rate at $x$ depends on the conditions
what we have around $\tilde x_1$ and $\tilde x_2$, i.e. deeper inside the interactive
matter. Consequently, the collision rate is still higher than the conditions
at $x$ could secure! Then  $ {\mathcal{P}}_f$ determines what fraction
of the outgoing particles will freeze out from those which collided.
The collision rate does not go to zero even if we are at the
outside boundary of the FO layer, because particles still can arrive
from the left where we still have interacting
particles. As there are no interacting particles on the right hand side,
all of these particles should freeze out, i.e. $ {\mathcal{P}}_f\rightarrow 1$,
when $x \rightarrow L$ (see Figure \ref{fig_mfp}).

Let us execute two of the phase space integrals for one incoming and one
outgoing particle,
$
\int \frac{d^3p_2}{p_2^0} \ \frac{d^3p_4}{p_4^0}\
$
\beq
p^\mu \partial_\mu f^f=
\frac{1}{2}\int {_{12}\mathcal{D}}_4  f^i_1 f^i_2 \ \
                     {\mathcal{P}}_f W_{12}^{p4}  =
\eeq{BTEhf-3}
\beq
 =
\frac{1}{2} Q_2 V_4 \int \frac{d^3p_1}{p_1^0}  f^i(\tilde x_1,p_1)  \ \
               {\mathcal{P}}_f W_{1\bar{2}}^{p\bar{4}},
\eeq{BTEhf-4}
where
$
Q_2     = \int \frac{d^3p_2}{p_2^0}\  f^i(\tilde x_2,p_2)
$
and  $V_4$ are invariant scalars.
Eq. (\ref{BTEhf-4}) resembles eq. (3.27) in ref. \cite{LasBook}, but one of
the incoming particle distributions,  $f^i(\tilde x_2,p_2) $ is integrated out,
and leads to an integral quantity, $Q_2$. This can be approximated by the
invariant scalar density at $\bar{x}_2$, i.e.
$$
Q_2 \approx n_2(\bar{x}_2) \ .
$$
Here $V_4$ is not known directly, but can in principle be calculated
based on the distributions, $f^i(\tilde x_1,p_1)$ and  $f^i(\tilde x_2,p_2)$,
and the transition rate, $W$.  The resulting transition rate
will then be averaged over particles 2 and 4, $W_{1\bar{2}}^{p\bar{4}}$:
\beq
p^\mu \partial_\mu f^f(x,p)=
\frac{1}{2} Q_2 V_4 \int \frac{d^3p_1}{p_1^0}  f^i(\tilde x_1,p_1)
               {\mathcal{P}}_f W_{1\bar{2}}^{p\bar{4}} \ .
\eeq{BTEhf-5}
As we see this resulting equation is delocalized -  in a rapid dynamical process the distributions at $x$ and $\tilde x_1$
are not the same, as it was discussed above.

Now, eq. (\ref{BTEhf-5}) can be integrated either in the $\tilde x_1$ -space,
or in the $p_1$ -space, as the two are connected by the fact that a
particle should travel from $\tilde x_1$ to $x$ with momentum $p_1$.\ \
We should integrate over all $\tilde x_1$ points from where one can reach
$x$ in a collision time, $\tau(x,\vec{v}/v)$. This brings in information about the
local spatial gradient of the ST distribution function, as we discussed
it above.
The collision terms in the original BTE contain only local information,
which is assumed to be isotropic (or slowly changing), so it is neglected.

In addition the FO probability, ${\mathcal{P}}_f$, may include
integrated information about the FO process, e.g. the probability
not-to-collide with anything on the way out, reasonably should
depend on the integral number of interacting particles on the way
out.

For the sake of simplicity let us assume small angle scatterings, and the propagation
of a single particle
$W_{1\bar{2}}^{p\bar{4}} \approx $
$w_{\bar{2}}^{\bar{4}}\  \delta(p-p_1)$,
then
\beq
p^\mu \partial_\mu f^f(x,p)=
\frac{1}{2} Q_2 V_4 \
f^i(\tilde x_1, p)\
{\mathcal{P}}_f\ w_{\bar{2}}^{\bar{4}}\,.
\eeq{BTEhf-6}
The cumulative effect of all particles which can reach the ST point
$x$ in a collision time, leads to a change directed into the direction
given by $d\sigma^\mu$. The transition rate, $ w_{\bar{2}}^{\bar{4}} $
can be estimated as $\langle \sigma v_{rel} \rangle \sim p^\mu / p^0$, what yields to:
\beq
p^\mu \partial_\mu f^f(x,p) = f^i(\tilde x_1,p) \
\left\{ \frac{1}{2}
 Q_2 V_4 \ {\mathcal{P}}_f\ \ d\sigma^\mu p_\mu / p^0 \ .
\right\}
\eeq{MBTE6}

As we mentioned, the spatial variation of the phase-space distribution
cannot be neglected in rapid dynamical processes as the FO, and this
brings in a direction of the dominant change, $d\sigma^\mu$.

Let us now consider the FO situation, where we have a directed process
in a layer. The dominant change happens in the direction of the
normal of the FO hypersurface,
$d\sigma^\mu $ (where $d\sigma^\mu \ d\sigma_\mu = \pm 1$).
We can decompose the
4-vector, $p^\nu$ on the l.h.s. of the above equations into four orthogonal
directions:
\bea
p^\nu &=&
   (p^\mu d\sigma_\mu) d\sigma^\nu  +
   (p^\mu d\sigma_{1\mu}) d\sigma_1^\nu  \\ \nonumber
&+&   (p^\mu d\sigma_{2\mu}) d\sigma_2^\nu +
      (p^\mu d\sigma_{3\mu}) d\sigma_3^\nu  \ ,
\eea
where the 4-vectors, $d\sigma_1^\nu$,  $d\sigma_2^\nu$, and $d\sigma_3^\nu$,
are tangent to the hypersurface and orthogonal to the normal, $d\sigma^\nu$.
This leads to:
\bea
p^\nu \partial_\nu f(x,p) &=& \big[
   (p^\mu d\sigma_\mu) d\sigma^\nu  + (p^\mu d\sigma_{1\mu}) d\sigma_1^\nu \\ \nonumber
&+& (p^\mu d\sigma_{2\mu}) d\sigma_2^\nu
+(p^\mu d\sigma_{3\mu}) d\sigma_3^\nu  \big] \partial_\nu f(x,p) \ .
\eea
Here we assumed
that the change happens in the direction of the normal
and negligible along the hypersurface of the front, thus the last three
terms can be neglected:
$$
  p^\nu \partial_\nu f(x,p) \approx
 (p^\mu d\sigma_\mu)\ d\sigma^\nu \partial_\nu\ f(x,p) \ .
$$
Inserting the above equation into (\ref{MBTE6}) yields a kinetic equation
describing the directional derivative of the distribution function
in the direction of the dominant change, $d\sigma^\mu$ as
\beq
d\sigma^\mu \partial_\mu f^f(x,p) = f^i(\tilde x_1,p) \  P^*_{esc} \ ,
\eeq{MBTE7}
where the escape probability depends on the ST coordinates, on the
interacting part of the PS distribution, on the transfer properties and
the FO probability:
$
P^*_{esc} (x,p,f^i,d\sigma,w, {\mathcal{P}}_f)
$.
The $\tilde x_1$ in this case means $\tilde x_1=x-d\sigma^\mu \lambda_\mu$,
where $\lambda_\mu$ is a four-vector of m.f.p.

This $\tilde x_1$ in the argument of the distribution function on the r.h.s. of the eq. (\ref{MBTE7})
is extremely important. Certanly, we can repeat all the steps from eq. (\ref{BTEhf-3}) to eq. (\ref{MBTE7}) based on the BTE amd the result will be the same except for the delocalization of $f$ on the r.h.s. This $x-d\sigma^\mu \lambda_\mu$ dependence of the distribution function reflects the FO property which was discussed at the beginning of this section and illustrated in Fig.~\ref{fig_mfp} - the collision rate at some point $x$, and correspondingly the number of particles, which will freeze out after this collision, feels the properties of the matter deeper (by about one m.f.p.) inside the interacting matter.

The derivation above did neglect several details and features, however,
reflects the basic structure of ad hoc kinetic FO models
\cite{FO1,FO3-HIP,FO4,FO5}. In these models the infinitely long FO was studied, and therefore the delocalization of eq. (\ref{MBTE7}) was not so important. For the FO modeling in the finite layer \cite{Mo04a} this effect will cause a substantial difference making FO faster.

\section{Escape Probability}
\label{escape}

The escape probability in eq. (\ref{MBTE7}) can be estimated based on
fundamental physical principles, like it is done in the above mentioned
works. The approach can, nevertheless, be improved if we take into consideration
the origin of the above derivation, especially the requirement of
full covariance of the model and the requirement that the FO process
may point in any space-time direction. The first significant advances,
where these principles were applied, are presented in \cite{Cs04}.
It incorporates the achievements of recent years, by cutting negative contributions
in the FO density \cite{Bu96} and making the
FO direction dependent \cite{FO1}.  Here we just present briefly a direct
estimate for the escape probability \cite{magastalk1,magastalk2}.

The escape probability includes the FO probability, which separates
from among the outgoing (gain) particles, which fraction of them
is still colliding and which not. The probability
not-to-collide with anything on the way out, reasonably should
depend on the number of particles, which are in the way of a particle
moving outwards in the direction $\vec{p}/p$, across a FO layer
of estimated thickness $L$ (representing the fact that we have finite number
of particles on the way out to collide with \cite{FO3-HIP}). If we are in this FO layer and
progressed
from the beginning of the layer to a position $x^\mu$, there is still
$$\frac{L - x^\mu d\sigma_\mu}{\cos \Theta}$$ distance ahead of us, where $\Theta$ is an
angle between the normal vector
and $\vec{p}/p$. We assume then that
the FO probability is inversely proportional to some power of this
quantity \cite{magastalk1,magastalk2}. Thus
\be
   P^*_{esc} =
   \frac{1}{\lambda(\bar{x}_1)}
   \left( \frac{L}{L-x^\mu d\sigma_\mu} \right)^a \left(\cos \Theta\right)^a
   \Theta(p^\mu d\sigma_\mu)\,,
\label{esc1a}
\ee
where the power $a$ is influencing the FO profile across the front, and
the cut factor is eliminating negative contributions to FO.
In papers \cite{FO1,FO3-HIP,FO4,FO5} the authors have used $a=1$,
and modeled FO in an infinite layer, i.e. in $L\rightarrow \infty$ limit. Furthermore,
they were using
a constant characteristic length $\lambda$ instead of  $\lambda(\bar{x}_1)$:
\be
P^*_{esc} =\frac{\cos \Theta}{\lambda}\  \Theta(p^\mu d\sigma_\mu)\,.
\label{oldFP}
\ee
Comparing eqs. (\ref{esc1a}) and (\ref{oldFP}) one can see that now we
replace the constant characteristic length $\lambda$,
which was clearly oversimplifying the situation, with two factors.
The first is the collision rate, which is proportional with
$
\frac{1}{\lambda(\bar{x}_1)} \approx  \langle  n(\bar{x}_1) \sigma  \rangle \ ,
$
and this does not tend to zero even if we reach the outside
boundary of the FO layer, as this parameter is characteristic
to the interior region at $\bar{x}_1$.
The other is the generalized FO probability, which depends on the direction of the
outgoing particle and on the number of interacting particles left in the
way to collide with, i.e. $\propto \frac{L}{L-x}$,
where we have fixed $d\sigma^{\mu}=(0,1,0,0)$.
So, we have generalized eq. (\ref{oldFP}) by replacing
\be
\lambda \rightarrow \lambda'(x) =\lambda(\bar{x}_1)\frac{L-x}{L}\,.
\label{lambda}
\ee
Now the new characteristic length
$\lambda'(x)$
gradually decreases as FO proceeds and the number of interacting particles
becomes smaller and smaller, and goes to $0$ when the FO is finished,
as it was discussed in section \ref{secMBTE}.

The simple angular factor, $\cos \Theta$, maximizes the FO
probability of those particles, which propagate
in the direction closest to the normal of the layer, $ d\sigma_\mu$.
The quantities,
$\cos \Theta=p^x/| \vec{p} |$ for FO in $x$-direction and
$\cos \Theta=1$     for FO in $t$-direction,
are not Lorentz invariant. Therefore, to make our description
completely invariant we shall generalize it to
$\frac {p^\mu d\sigma_\mu}{p^\mu u_\mu} \sim \cos \Theta$.

So, we write the invariant escape probability, within the FO layer
covering both the time-like and space-like parts of the layer \cite{magastalk2},
as
\be
   P^*_{esc} =
   \frac{1}{\lambda(\bar{x}_1)}
   \left( \frac{L}{L-x^\mu d\sigma_\mu} \right)^a
   \left(\frac {p^\mu d\sigma_\mu}{p^\mu u_\mu}\right)^a\
   \Theta(p^\mu d\sigma_\mu)\,.
\label{esc1}
\ee

If we take the four velocity equal to $u_{\mu}=(1,0,0,0)$, in the
Rest Frame of the Front (RFF), i.e. where $d\sigma_{\mu} = (1,0,0,0)$,  then
the momentum dependent part of the Escape Probability, $P(p)$,
is unity. Otherwise, in the Rest Frame of the Gas (RFG),
where $u_{\mu}=(1,0,0,0)$, the escape probability $P(p)$ is
$
    P(p) =
    {p^{\mu}d\sigma_{\mu}} \
    \Theta(p^{\mu}d\sigma_{\mu})/{p^{0}}.
%    \label{esc2}
$
More detailed investigation about escape probability $P(p)$ for different $d\sigma_{\mu}$ can be found in \cite{Cs04}.

To calculate the parameters of the normal vector $d\sigma_{\mu}$
for different cases listed above, we simply make use
of the Lorentz transformation. The normal vector
of the time-like part of the FO hypersurface may be
defined as the local $t'$-axis, while the normal vector for
the space-like part may be defined as the local $x'$-axis.
As the $d{\sigma}_\mu$ normal vector is normalized to unity
its components may be interpreted in terms of
$\gamma_{\sigma}$ and $v_{\sigma}$, as
$d\sigma_{\mu}=\gamma_{\sigma} (1,v_{\sigma},0,0)$,
where
$\gamma_{\sigma} = \frac{1}{\sqrt{1 - v_{\sigma}^2}}$
for time-like normals and
$\gamma_{\sigma} = \frac{1}{\sqrt{v_{\sigma}^2-1}}$
for space-like normals.

The detailed results of the application of this covariant
escape probability will be presented elsewhere
\cite{Mo04a}.

In refs. \cite{FO3-HIP,FO4} the post FO distribution was evaluated for
space-like gradual FO in a kinetic model. Initially we had an equilibrated,
interacting PS distribution, $f_{int}(p,x)$, and an escape probability,
similar to eq. (\ref{esc1}), but simplified one. It was dependent on the
angle of the two vectors only. After some small fraction of particles
were frozen out as the FO process progressed in the front, the interacting
component were re-equilibrated with smaller particle number, smaller
energy and momentum to account for the quantities carried away by the
frozen out particles. This was then repeated many times in small steps
along the FO front and the frozen out particles were accumulated in the
post FO PS distribution, $f_{free}$. The resulting distribution
was highly anisotropic and obviously non-equilibrated. The details of the
post FO distribution depend on the details of the escape probability, and
on the level of re-equilibration of the remaining, interacting component.

Bugaev assumed earlier \cite{Bu96} (see also ref. \cite{E}), that the post FO distribution is a
(sharply cut) "Cut-J\"uttner" distribution, but the above mentioned model shows
that this can only be obtained if re-equilibration is not taking place.
The kinetic model provided an asymmetric but smooth PS distribution
\cite{FO3-HIP}, while the escape probability (\ref{esc1}) yields a somewhat
different, but also smooth PS distribution \cite{magastalk2}. These can be well approximated
by the "Canceling J\"uttner" distribution \cite{Ta03}.

In ref. \cite{FO5} the same infinite 1D model (as in refs. \cite{FO3-HIP,FO4})
was applied for the FO through the layer with
time-like normal vector. The model in this case can be solved analytically (since $\cos
\Theta \equiv 1$) and thus, the exact form of the post FO distribution for pions and protons
was obtained. Although analytical expressions for these distribution are different from the thermal
J\"uttner distribution, the forms of the functions are very similar for intermediate and high momenta.
Deviations at the low momenta seems to be due to infinite long FO (they become much smaller for
the escape probability (\ref{esc1}), modeling FO in a finite layer) and 1D character of the model
\cite{magastalk2}.

\section{Conclusions}

The FO process was discussed in the 4-di\-men\-si\-onal
space-time in a layer of finite thickness.  Arising from the
physical process this layer is directed, it has an inside and outside
boundary, which are not identical. The processes in the layer
are not isotropic, they must be sensitive to the direction of the
layer.  It is shown that, as a consequence, the basic assumptions of
the Boltzmann Transport Equation are not satisfied in this layer,
and the equation should be modified. It is also shown that earlier,
ad hoc kinetic models of the FO process, can be obtained
from this approach in a fully covariant way, and freeze out
in space-like and time-like directions can be handled on the same
covariant footing.

\bigskip

{\it Acknowledgments:\ \ }
One of the authors,  L.P. Cs., thanks the Alexander von Humboldt Foundation
for extended support in continuation of his earlier Research Award.
The authors thank the hospitality of the Frankfurt Institute of Advanced
Studies and the Institute for Theoretical Physics of the University of
Frankfurt, and the Gesellschaft f\"ur Schwerionenforschung, where parts
of this work were done.
Enlightening discussions, with Cs. Anderlik, K.A. Bugaev, F. Grassi, Y. Hama,
T. Kodama and D. Strottman are gratefully acknowledged.

\vfill
\end{document}